\documentclass[twocolumn,twoside,10pt]{IEEEtran}
\usepackage{setspace}
\usepackage{cite,color}%
\usepackage{amsfonts}%
\usepackage{amsmath,url}%
\usepackage{amssymb}%
\usepackage{graphicx}
\usepackage{balance,epsfig}
\usepackage{amsmath,amssymb,graphicx,epsfig,url,cite,balance}
\begin{document}
\newcounter{eqncount}
\pagestyle{empty}
\thispagestyle{empty}

\title{Performance of Cognitive Radio Systems with Imperfect Radio Environment Map Information}
\author{\authorblockA{Muhammad Fainan Hanif~\IEEEauthorrefmark{1}, Peter J.
Smith~\IEEEauthorrefmark{1} and Mansoor
Shafi~\IEEEauthorrefmark{2}}\\
\authorblockA{\IEEEauthorrefmark{1}Department of Electrical and Computer Engineering, University of Canterbury,
Christchurch, New Zealand}
\authorblockA{\IEEEauthorrefmark{2}~Telecom New Zealand, PO Box 293, Wellington, New Zealand\\
Email:mfh21@student.canterbury.ac.nz,~p.smith@elec.canterbury.ac.nz,~mansoor.shafi@telecom.co.nz}}
\maketitle \thispagestyle{empty} \pagestyle{empty}
\begin{abstract}
In this paper we describe the effect of imperfections in the radio
environment map (REM) information on the performance of cognitive
radio (CR) systems. Via simulations we explore the relationship
between the required precision of the REM and various channel/system
properties. For example, the degree of spatial correlation in the
shadow fading is a key factor as is the interference constraint
employed by the primary user. Based on the CR interferers obtained
from the simulations, we characterize the temporal behavior of such
systems by computing the level crossing rates (LCRs) of the
cumulative interference represented by these CRs. This evaluates the
effect of short term fluctuations above acceptable interference
levels due to the fast fading. We derive analytical formulae for the
LCRs in Rayleigh and Rician fast fading conditions. The analytical
results are verified by Monte Carlo simulations.
\end{abstract}
\section{Introduction}
It is now well known \cite{report1,report2} that granting exclusive
licences to service providers for particular frequency bands has led
to severe under-utilization of the radio frequency (RF) spectrum.
This has led to global interest in the concept of cognitive radios
(CRs) or secondary users (SUs). These CRs are deemed to be
intelligent agents capable of making opportunistic use of radio
spectrum while simultaneously existing with the legacy primary users
(PUs) without harming their operation.

In addition to ensuring quality of service (QoS) operation, the most
important and challenging task for the CRs is to avoid adverse
interference to the incumbent PUs. Hence, it is necessary to develop
schemes that can help PUs avoid such harmful interference. In
addition to the \emph{primary exclusion zone} (PEZ) \cite{Mai1}
approach, the recently developed \cite{ICC} methods based on radio
environment maps (REMs) \cite{J.Reed1,J.Reed2} can also help achieve
this goal. In an earlier paper \cite{ICC} we have shown that under
certain conditions the REM based approach can result in
substantially higher numbers of permissible CRs than the PEZ
approach. Hence, the focus of this paper is an REM scheme. In
particular we consider a REM which stores signal strength data from
point to point in a regular grid. CRs have access to this REM and
can therefore evaluate their impact on the PU and maintain
acceptable interference levels as long as they can obtain positional
information on the other CRs and the PU.

The REM based approaches heavily depend on ``quantity'' and
``quality'' of the REM information available. Defects in REM
information can seriously affect the PU performance. In a similar
manner, temporal variations in the CRs' interfering signals can
degrade the PU performance even though the CR level may be
acceptable on average. These two aspects form the focus of this
paper. In both situations we assume that the PU is willing to suffer
some reduction in SNR, so that an allowable level of interference is
provided to enable the CR operation. In particular, we make the
following contributions:
\begin{itemize}
\item We determine the impact of coarse REM information. We show
that when the REM for a given area is discretized  then the total CR
interference is significantly underestimated when realistic grid
sizes are considered. For example, for a grid size of 50 m $\times$
50 m, the actual SINR is worse than the target SINR by at least 1 dB
for 8\% of the time. We also determine the interaction between
shadow fading correlation and REM grid size and evaluate their
impact on interference estimation.
\item We determine the level crossing rate (LCR) and exceedance
duration (AED) of the CR-PU interference for a number of scenarios
including Rayleigh fading, Rician fading, and various CR interferer
profiles. The LCR is determined via analysis and confirmed via
simulation. Results show that the LCR is maximum at or around the
maximum interference threshold and is virtually zero 5 dB beyond
this point. We also show that for urban areas which are
characterized by a strong LOS component, the interference rarely
crosses the threshold and when it does, it only exceeds the
threshold value for small duration.
\end{itemize}
The rest of the paper is organized as follows: Section~II describes
the system model and the REM. Section~III characterizes the
instantaneous composite CR interference to the PU system in terms of
the LCRs. In Section~IV we present simulation and analytical
results. Finally, in Section~V we describe our conclusions.
\section{System Model and REM}
Consider a PU receiver in the center of a circular region of radius
$R$. The PU transmitter is located uniformly in an annulus of outer
radius $R$ and inner radius $R_{0}$ centered on the PU receiver. It
is to be noted that we place the PU receiver at the center only for
the sake of mathematical convenience. The use of the annulus
restricts devices from being too close to the receiver. This matches
physical reality and also avoids problems with the classical inverse
power law relationship between signal strength and distance
\cite{Mai}. In particular, having a minimum distance, $R_0$,
prevents the signal strength from becoming infinite as the
transmitter approaches the receiver. Similarly, we assume that
multiple CR transmitters are uniformly located in the annulus. At
any given time, each CR has a probability of seeking a connection,
given by the activity factor, $p$. The number of CRs wishing to
operate is denoted $N_{CR}$. Of these CRs, a certain number will be
accepted depending on the allocation mechanism. Hence, a random
number of CRs denoted $N\leq N_{CR}$ will transmit during the PU
transmission and create interference at the PU receiver.

The received signal strength for both the PU transmitter to PU
receiver and CR transmitter to PU receiver is assumed to follow the
classical distance dependent, lognormal shadowing model. For a
generic interferer, this is given by
\begin{equation}\label{first}
I=BLr^{-\gamma}=B10^{\tilde{X}/10}r^{-\gamma}= Be^Xr^{-\gamma}
\end{equation}
where $r$ is the random distance from the transmitter to the
receiver, $\gamma$ is the path loss exponent (normally in the range
of 2 to 4) and $L$ is a shadow fading variable. The lognormal
variable, $L$, is given in terms of the zero mean Gaussian,
$\tilde{X}$, which has standard deviation $\sigma$ (dB) and
$X=\beta\tilde{X}$ where $\beta=\log(10)/10$. The standard deviation
of $X$ is denoted by $\sigma_{x}$. The constant $B$ is determined by
the transmit power. The desired primary signal strength, $S$, has
the same form, with a different transmit power, so that
$S=AL_pr_p^{-\gamma}$. The constant $A$ is determined so as to give
an SNR greater than 5 dB, 95\% of the time. However, the constant
$B$ depends upon $A$, $\gamma$ and the ratio of $R$ and $R_c$ as
given in \cite{ICC}. Note that all the links are assumed to be
independent and identically distributed (i.i.d.) so that spatial
correlation is ignored.
\subsection{A Perfect REM}
A REM can hold a wide variety of information \cite{J.Reed2} and it
is not clearly understood at present what constitutes a practical
and effective REM. In this work we assume that the REM contains
signal strength data. In a perfect REM the signal strength from all
source coordinates to all destination coordinates is known. With
this perfect REM a CR controller \cite{ICC} can select those CRs for
operation which satisfy a given interference constraint. The CR
controller requires positional information for the PU and the CRs,
and can then use the REM to compute the overall SINR of the PU where
\begin{equation}\label{sinr}
\textrm{SINR}=\frac{S}{\sum_{i=1}^NI_i+\sigma^2}
\end{equation}
In (\ref{sinr}), $S$ is the signal strength of the PU, $\sigma^2$ is
the noise power and $\sum_{i=1}^NI_i$ is the aggregate interference
of the $N$ selected CRs. The interference constraint used is that
the CR interference must not reduce the PU SNR by more than 2 dB.
All results shown in the paper are for 2 dB buffer. The value of 2
dB was chosen arbitrarily and an exploration of effects of this
buffer will appear in a future work. In this paper the centralized
approach in \cite{ICC} is used to select the $N$ CRs for operation.
\subsection{Modeling of REM Imperfections}
In practise a perfect REM is impossible and for practical purposes
the REM information is discretized in the form of grid points with
grid size, $\Delta$. Hence, the central controller allocating CRs
will formulate its decisions on the basis of REM information
obtained from the grid points, rather than from exact signal
strength data. Hence, an interfering signal strength, $I$, will be
estimated by $\hat{I}$ from the REM. The estimate is obtained from
the grid-to-grid path in the REM which is closest to the actual
signal path.

We consider the CR signal strength to be of the form given in
(\ref{first}). The REM predicted signal strength is given by:
\begin{equation}\label{first1}
\hat{I}=Be^{\hat{X}}\hat{r}^{-\gamma}
\end{equation}
where $\hat{r}$ is the distance between the transmitter and the
receiver in the REM grid and $\hat{X}$ is correlated with $X$ by:
\begin{equation}\label{gans}
\hat{X}=\rho X+\sqrt{1-\rho^2}E
\end{equation}
In (\ref{gans}) $E$ is independent and identically distributed
(i.i.d.) with $X$. Assuming a distance, $d_i$, between the actual
and REM based position of the CR and a distance, $d_p$, between the
actual and REM based location of the PU receiver, the correlation
coefficient $\rho$ can be obtained using an extension of
Gudmundson's model \cite{Gud} as:
\begin{equation}\label{model}
\rho=0.5^{d_i/D_d}\times 0.5^{d_p/D_d}
\end{equation}
In (\ref{model}) $D_d$ is the so called \emph{decorrelation
distance} i.e., the distance at which the correlation between $X$
and $\hat{X}$ drops to 0.5. The effect of flawed REM information on
the signal strength between the primary transmitter and its receiver
can also be modeled using (\ref{first1}), (\ref{gans}) and
(\ref{model}). Simulation results of this model based on parameter
values of a suburban macrocellular environment are given in
Section~IV.
\section{Instantaneous CR performance}
The CR allocation policy is based on mean signal and interference
levels. As a result, even if the 2 dB buffer is exactly met the
instantaneous fast fading will result in fluctuations of the SINR
both above and below the buffer. It is therefore of interest to
investigate how often and how long the SINR exceeds the buffer. As a
first look at this problem we fix the PU signal power and consider
the instantaneous variation of the interference only. In this
scenario the 2 dB SINR buffer becomes a threshold of -2.33 dB for
the interference (as shown in Figs. 4-7). Hence in this section we
focus on the instantaneous temporal behavior of the aggregate
interference. For this purpose we evaluate the LCR (and thus the
average exceedance duration (AED)) of the cumulative interference
offered by the CRs obtained using the \emph{centralized approach} of
\cite{ICC} with imperfect REM information. First we calculate the
LCRs for Rayleigh environment and then we characterize them for
Rician fading conditions. In future work the full temporal behavior
of the SINR should be considered, but this preliminary investigation
still yields useful results and insights.
\subsection{LCRs for Rayleigh Fading}
For a given set of CR interferers, the instantaneous aggregate
interference under Rayleigh fading, $I_{Ray}(t)$, is given by:
\begin{equation}\label{instRay}
I_{Ray}(t)=\sum_{i=1}^NI_i|h_i(t)|^2
\end{equation}
where $I_i$ represents the interference power of the $i^{th}$ CR,
$|h_i(t)|^2$ is a standard exponential random variable with unit
mean and $N$ is the number of interfering CRs. From (\ref{instRay}),
the aggregate interference is represented as a weighted sum of
exponential random variables. Such weighted sums can be approximated
by a gamma variable \cite{Kutz}. Simulated results show that the
gamma fit is very good, but are not shown here for reasons of space.
It should be noted that the exact LCR computation for such weighted
sums was given in \cite{MRC2} for the case of three and four branch
maximal ratio combining (MRC) by providing special function
integrals. Recently, more general expressions for arbitrary number
of branches have been derived in \cite{MRC1}. However, the approach
of \cite{MRC1} results in numerical difficulties, especially for
large values of $N$, which can be the case for CR systems. Hence an
approximation is useful to overcome these problems and to provide a
much simpler solution. A gamma variable with shape parameter $r$ and
scale parameter $\theta$ has a mean and variance given by $r/\theta$
and $r/\theta^2$ respectively and probability density function
(PDF):
\begin{equation}\label{gammaRV}
f(x)=\Gamma(r)^{-1}\theta^rx^{r-1}\exp(-\theta x),\qquad x\geq0
\end{equation}
Thus, approximate LCRs for (\ref{instRay}) can be found by
calculating the LCR of the equivalent gamma process.
The LCR for a gamma process has been calculated in \cite{lcrGamma}.
Thus, the crossing rate of $I_{Ray}(t)$ across a threshold, $T$, can
be approximated by:
\begin{equation}\label{lcrRay}
\textrm{LCR}_{I_{Ray}}(T)=\frac{1}{2\Gamma(r)}\sqrt{\frac{2|\ddot{R}(0)|}{\pi}}(\theta
T)^{r-0.5}\exp(-\theta T)
\end{equation}
where $r=E(I_{Ray}(t))^2/Var(I_{Ray}(t))$,
$\theta=E(I_{Ray}(t))/Var(I_{Ray}(t))$ and
$\ddot{R}(0)=\ddot{\rho}_{Ray}(0)$ is the second derivative of the
autocorrelation function (ACF) of $I_{Ray}(t)$. Hence, to compute
the LCR in (\ref{lcrRay}) only the mean, variance and ACF of the
random process in (\ref{instRay}) are required.

The first two moments of (\ref{instRay}) are simple to compute as
$E(I_{Ray}(t))=\sum_{i=1}^NI_i$ and
$Var(I_{Ray}(t))=\sum_{i=1}^NI_i^2$. To calculate the ACF, note that
$h_i(t+\tau)$ can be written as:
\begin{equation}\label{corrEq}
h_i(t+\tau)=\rho_i(\tau)h_i(t)+\sqrt{(1-\rho_i^2(\tau))}e_i(t),
\end{equation}
where $e_i(t)$ is independent of $h_i(t)$ and statistically
identical to $h_i(t)$. Assuming a Jakes' fading process,
$\rho_i(\tau)$ is the zeroth order Bessel function of the first
kind, $J_0(2\pi f_D\tau)$ and $f_D$ is the Doppler frequency. Using
(\ref{corrEq}) we have:
\setlength{\arraycolsep}{0.0em}
\begin{eqnarray}\label{meantT}
E[I_{Ray}(t)I_{Ray}(t+\tau)]&{}={}&\sum_{i,j=1}^NI_iI_jE[|h_i(t)|^2|h_j(t+\tau)|^2]\nonumber\\
&{}={}&\sum_{i\neq
j}^NI_iI_j+\bigg(\sum_{i=1}^NI_i^2E[|h_i(t)|^2(\rho_i^2(\tau)\nonumber\\
&&{\times}\:|h_i(t)|^2+(1-\rho_i^2(\tau))|e_i(t)|^2)]\bigg)\nonumber\\
&{}={}&\sum_{i\neq j}^NI_iI_j+\sum_{i=1}^NI_i^2+\sum_{i=1}^NI_i^2\rho_i^2(\tau)\nonumber\\
&{}={}&\bigg(\sum_{i=1}^NI_i\bigg)^2+\sum_{i=1}^NI_i^2\rho_i^2(\tau),
\end{eqnarray}
\setlength{\arraycolsep}{5pt}
\hspace{-1.5mm}where in the second to last step above, we have used
the fact that cross products have zero mean and that
$E[|h_i(t)|^4]=2$. The ACF of (\ref{instRay}) is given by:
\begin{equation}\label{acf}
\rho_{Ray}(\tau)\!=\!\frac{E(I_{Ray}(t)I_{Ray}(t+\tau))\!-\!E(I_{Ray}(t))E(I_{Ray}(t+\tau))}{\sqrt{Var(I_{Ray}(t))Var(I_{Ray}(t+\tau))}},
\end{equation}
and with the relevant substitutions, the ACF becomes:
\begin{equation}\label{trueAcf}
\rho_{Ray}(\tau)=\frac{\sum_{i=1}^NI_i^2J_0^2(2\pi
f_D\tau)}{\sum_{i=1}^NI_i^2}.
\end{equation}
Finally, using the expansion $J_0(2\pi
f_D\tau)=1-\pi^2f_D^2\tau^2+\ldots$, the second derivative of the
ACF needed to compute the LCR in (\ref{lcrRay}) is evaluated as:
\begin{equation}\label{dder}
\ddot{\rho}_{Ray}(0)=-4\pi^2\frac{\sum_{i=1}^NI_i^2f_D^2}{\sum_{i=1}^NI_i^2}.
\end{equation}
Hence the three parameters, $r$, $\theta$ and $\ddot{R}(0)$, are
available and (\ref{lcrRay}) gives the approximate LCR.
\subsection{LCRs for Rician Fading}
As in the Rayleigh fading case, the instantaneous aggregate
interference, $I_{Ric}(t)$, for this scenario is given as:
\begin{equation}\label{instRic}
I_{Ric}(t)=\sum_{i=1}^NI_i|h_i(t)|^2,
\end{equation}
where $h_i(t)$ is Rician and $N,I_1,I_2,\ldots,I_N$ are as defined
in (\ref{instRay}). Hence, $I_{Ric}(t)$ is a weighted sum of
noncentral chi-square ($\chi^2$) random variables. Using the same
approximation philosophy as that used in the Rayleigh case, we
propose approximating (\ref{instRic}) by a single non-central
$\chi^2$. This approach is less well documented but has appeared in
the literature (see \cite{Kim}). Also note that a scaled, rather
than a standard, noncentral $\chi^2$ distribution is required for
fitting. A noncentral $\chi^2$ variable with $v$ degrees of freedom,
non-centrality parameter $\lambda$ and scale parameter $\alpha$ has
the following PDF:
\begin{equation}\label{ncx2}
p(x)=\frac{\alpha}{2}\exp\bigg(\frac{-(\lambda+\alpha
x)}{2}\bigg)\bigg(\frac{\alpha
x}{\lambda}\bigg)^{\frac{v-2}{4}}I_{\frac{v-2}{2}}\big(\sqrt{\lambda
\alpha x}\big),
\end{equation}
where $I_{(v-2)/2}$ is a modified Bessel function of the first kind
with order $(v-2)/2$. Fitting the PDF in (\ref{ncx2}) to the
variable in (\ref{instRic}) is performed using the method of moments
technique so that the approximate noncentral $\chi^2$ has the same
first three moments as $I_{Ric}(t)$. Note that there can be
numerical difficulties with the approach for certain values of
$I_1,I_2,\ldots,I_N$. Results are shown in Sec.~IV for cases where
the estimation procedure was successful. Further research is
necessary to make the methodology robust to all possible
interference values.

Next we consider the LCR of the noncentral $\chi^2$ process which is
used to model $I_{Ric}(t)$. Consider a generic scaled noncentral
$\chi^2$ process, $g$, given by:
\begin{equation}\label{ratio}
g=\frac{\sum_{i=1}^v(X_i+\delta_i)^2}{\alpha},
\end{equation}
where $\sum_{i=1}^v\delta_i^2=\lambda$ is the non-centrality
parameter, $v$ is the order (degrees of freedom) and $\alpha$ is the
scale parameter. Note that for simplicity we have omitted the
dependence on time so that $g(t)$ is denoted by $g$. The variables
$X_1,X_2,\ldots,X_v$ are i.i.d. $\mathcal{N}(0,1)$. Using the basic
formula of Rice, the LCR of $g$ across a threshold, $T$, is given
by:
\setlength{\arraycolsep}{0.0em}
\begin{eqnarray}\label{rice}
\textrm{LCR}_{I_{Ric}}&{}={}&\int_{0}^{\infty}\dot{g}p_{g,\dot{g}}(T,\dot{g})d\dot{g}\nonumber\\
&{}={}&\int_0^{\infty}\dot{g}p_{\dot{g}|g}(\dot{g}|T)p_g(T)d\dot{g},
\end{eqnarray}
\setlength{\arraycolsep}{5pt}
\hspace{-1.9mm}where $p_g(.)$, $p_{g,\dot{g}}(.,.)$ and
$p_{\dot{g}|g}(.)$ are the PDF of $g$, the joint PDF of $g$ and
$\dot{g}$ and the conditional PDF of $\dot{g}$ given $g$
respectively. The derivative of (\ref{ratio}) with respect to time
gives:
\begin{equation}\label{derratio}
\dot{g}=\frac{\sum_{i=1}^v2(X_i+\delta_i)\dot{X_i}}{\alpha}
\end{equation}
Now, using (\ref{ratio}) and (\ref{derratio}),
$p_{\dot{g}|g}(\dot{g}|T)$ is a Gaussian PDF corresponding to the
distribution $\mathcal{N}(0,4\gamma T/\alpha)$ with
$\gamma=Var(\dot{X})$. Thus, the LCR in (\ref{rice}) becomes:
\setlength{\arraycolsep}{0.0em}
\begin{eqnarray}\label{lcrRic}
\textrm{LCR}_{I_{Ric}}&{}={}&\frac{p_g(T)\sqrt{\alpha}}{\sqrt{8\pi\gamma T}}\int_{0}^{\infty}\dot{g}\exp\Bigg(\frac{-\alpha\dot{g}^2}{8\gamma T}\Bigg)d\dot{g}\nonumber\\
&{}={}&p_g(T)\sqrt{\frac{2\gamma T}{\alpha\pi}}
\end{eqnarray}
\setlength{\arraycolsep}{5pt}
Assuming the Jakes' fading process for the Gaussian variables,
$X_1,X_2,\ldots,X_M$, each $X_i$ has ACF given by $J_0(2\pi
f_D\tau)=1-\pi^2 f_D^2\tau^2+\ldots$. Hence,
$\gamma=Var(\dot{X})\rightarrow 2\pi^2 f_D^2$ as $\tau\rightarrow0$
\cite{Abromowitz}. Substituting, $p_g(T)$ from (\ref{ncx2}), the LCR
becomes,
\setlength{\arraycolsep}{0.0em}
\begin{eqnarray}\label{lcrRic1}
\textrm{LCR}_{I_{Ric}}&{}={}&p_g(T)\sqrt{\frac{4\pi^2f_D^2T}{\alpha\pi}}\nonumber\\
&{}={}&\sqrt{\pi}f_D(\alpha
T)^{\frac{v}{4}}\lambda^{\frac{-(v-2)}{4}}e^{\big(\frac{-\lambda-\alpha
T}{2}\big)}I_{\frac{v-2}{2}}\big(\sqrt{\lambda\alpha T}\big).\nonumber\\
\end{eqnarray}
\setlength{\arraycolsep}{5pt}
Note that this is the LCR of the process in (\ref{ratio}), and
appears to be a new result. However, the noncentral $\chi^2$ which
fits $I_{Ric}(t)$ will almost certainly not have an integer order.
Hence, this derivation for integer ordered noncentral $\chi^2$ is
applied to the fractional order case of interest. At present the
validity of this approach is only a conjecture, but the results in
Sec.~IV are very encouraging. Note that a similar extension for a
central $\chi^2$ with integer order \cite{chiP} to a central
$\chi^2$ with fractional order \cite{lcrGamma} has been shown to be
correct. A comparison of the theoretical results given in
(\ref{lcrRay}) and (\ref{lcrRic1}) with Monte Carlo simulations
along with a commentary on the results is given in the next section.
\section{Results}
\begin{figure}[t]
\centering
\includegraphics[width=0.95\columnwidth]{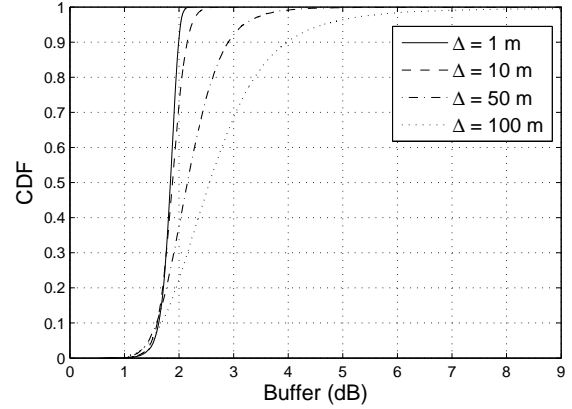}
\caption{Interference CDF for an REM enabled CR network for several
values of $\Delta$ and decorrelation distance, $D_d=100$ m.}
\label{fig_1} \vspace{-3mm}
\end{figure}
\begin{figure}[t]
\centering
\includegraphics[width=0.95\columnwidth]{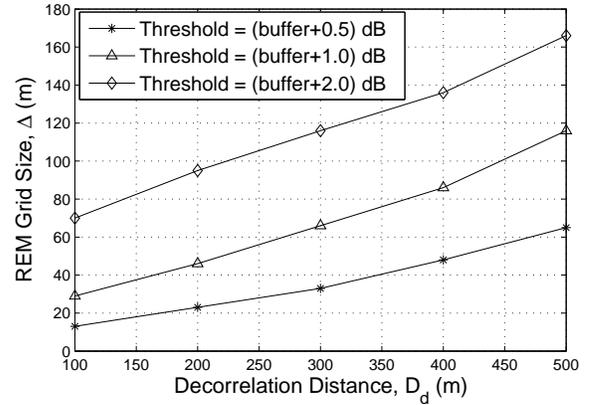}
\caption{REM grid size, $\Delta$, vs decorrelation distance, $D_d$,
for different threshold buffer sizes.} \label{fig_2}
\end{figure}
\begin{figure}[t]
\centering
\includegraphics[width=0.97\columnwidth]{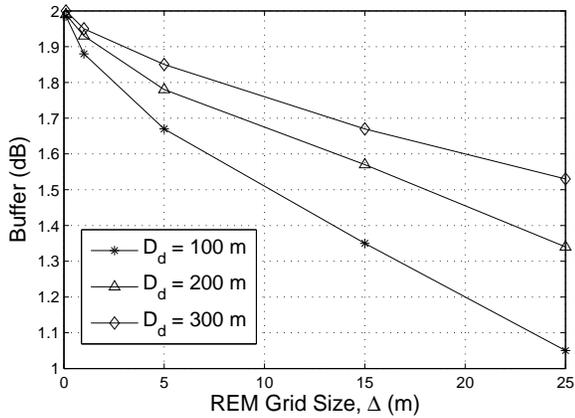}
\caption{Variation of actual buffer with REM grid size, $\Delta$,
for different values of the decorrelation distance, $D_d$.}
\label{fig1}
\end{figure}
\begin{figure}[t]
\centering
\includegraphics[width=0.95\columnwidth]{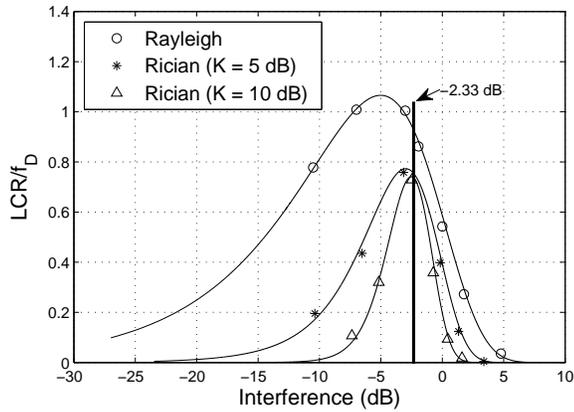}
\caption{LCR results for different fading conditions. The solid
lines represent analytical results. Simulation values are shown by
the circle, star and triangle symbols.} \label{fig2}
\end{figure}
\begin{figure}[t]
\centering
\includegraphics[width=0.95\columnwidth]{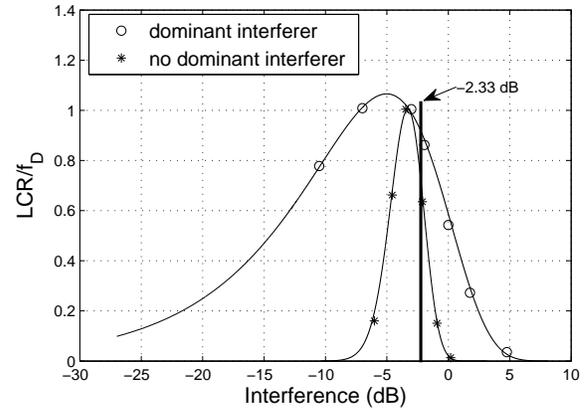}
\caption{LCR results for the dominant and no dominant interferer
cases in a Rayleigh fading scenario. The solid lines represent
analytical results. Simulation values are shown by the circle and
star symbols.} \label{fig3}
\end{figure}
\begin{figure}[t]
\vspace{0.0mm} \centering
\includegraphics[width=0.95\columnwidth]{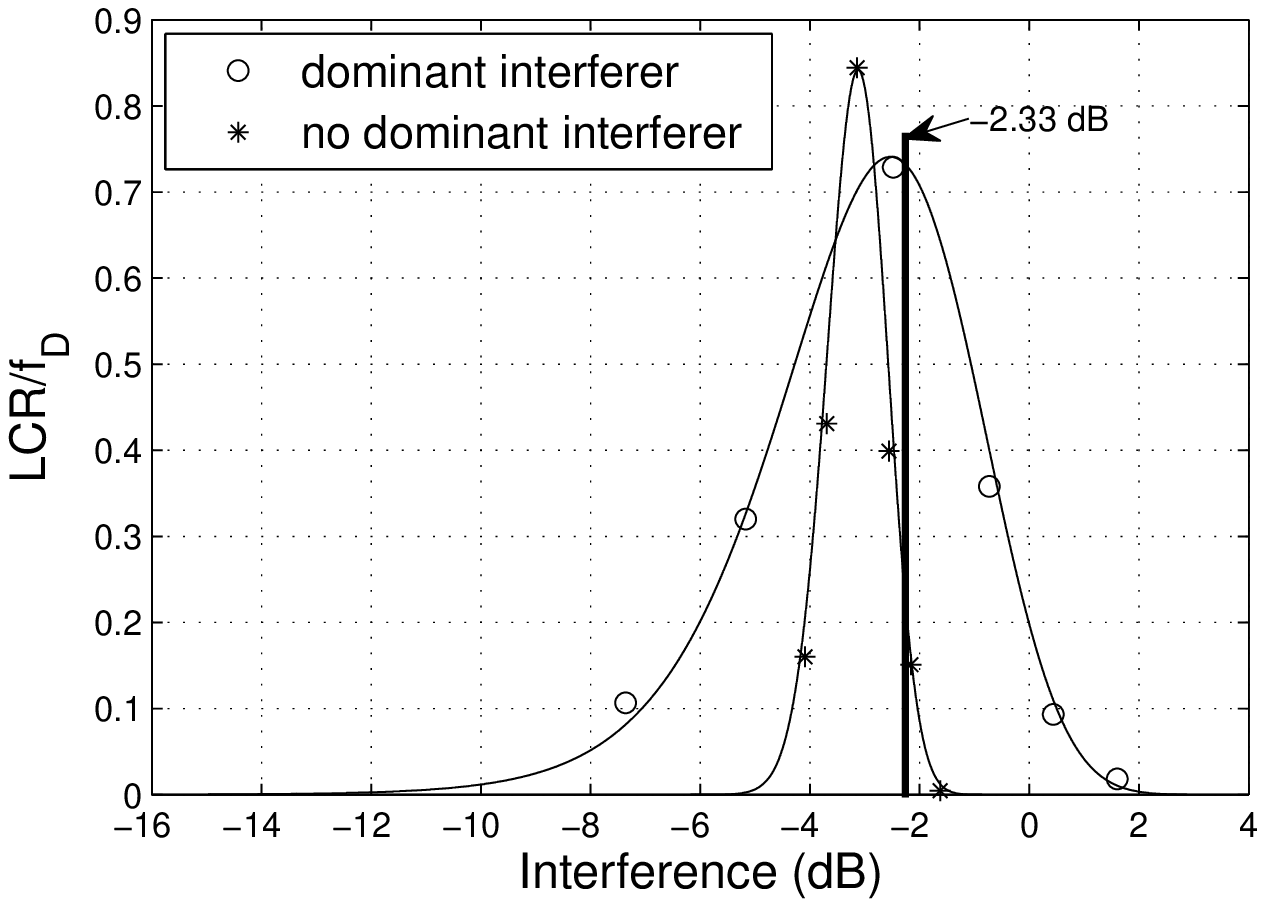}
\caption{LCR results for the dominant and no dominant interferer
cases in a Rician ($K=10$ dB) fading scenario. The solid lines
represent analytical results. Simulation values are shown by the
circle and star symbols.} \label{fig4}
\end{figure}
\begin{figure}[t]
\centering
\includegraphics[width=0.95\columnwidth]{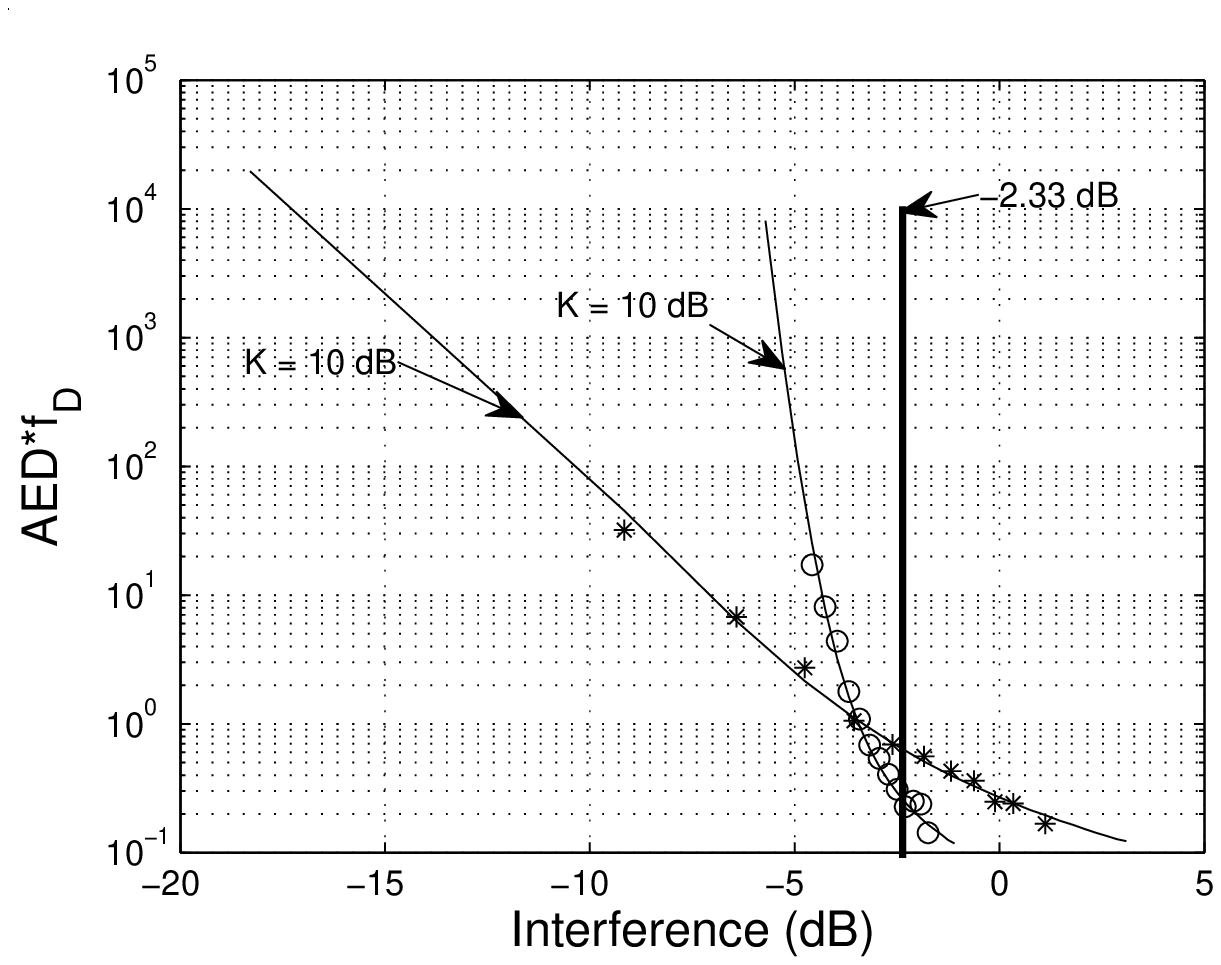}
\caption{AED results for the dominant and no dominant interferer
cases in a Rician ($K=10$ dB) fading scenario. The solid lines
represent analytical results. Simulation values are shown by the
circle and star symbols.} \label{fig_1} \vspace{-3mm}
\end{figure}
Throughout the section we assume the following parameter values:
shadow fading variance, $\sigma=8$ dB, path loss exponent,
$\gamma=3.5$, radius of PU coverage area, $R=1000$ m, radius of CR
coverage area, $R_c=100$ m, CR density, 1000 CRs per square
kilometer, an activity factor of 0.1 and $f_D=25$ Hz.
\subsection{Imperfections in the REM}
In practice, the  radio environment is often modeled by dividing an
area into a  regular grid (typically composed of 100 m $\times$ 100
m grid boxes) and assuming that the fading conditions in any grid
box can be approximated by a single point at the center of the box.
For example, drive testing of cellular networks to validate path
loss models and predicted signal coverage follows this approach.
Clearly, larger grid sizes result in errors between measurement and
prediction. On the other hand, reducing the grid size results in a
large data overhead. Figure 1 shows the cumulative distribution
function (CDF) of the magnitude of the actual CR-PU interference
when the REM is estimated via a grid size ranging from 1 m $\times$
1 m to 100 m $\times$ 100 m. The REM approach aims to maintain a 2
dB SINR buffer for the primary, but this is only possible with a
perfect REM. When $\Delta=1$ m the 2 dB buffer is nearly achieved
but for a grid size of 50 m $\times$ 50 m, the interference exceeds
3 dB for approximately 8\% of the time. For a grid size of 100 m
$\times$ 100 m, 3 dB is exceeded 30\% of the time. In effect this
means that if REM information is derived from a coarse grid, the
buffer size must be increased or the CRs must back off from the
buffer.

The effects of increasing the buffer or backing off the CRs are
shown in Figs.~2 and 3 respectively. In Fig.~2 the PU has a target 2
dB buffer but due to the imperfect REM it will not always be
achieved. Hence an extra buffer is permitted beyond which the CRs
are only allowed, 5\% of the time. In Fig.~2 this scenario is
denoted by the legend, Threshold = (original + extra) dB. The
effects of spatially correlated shadow fading are also considered in
Fig.~2. Shadow fading is correlated over any given area and the
level of this correlation has a simple effect on the REM grid size.
For highly correlated areas a coarse grid (large $\Delta$) will be
acceptable whereas in areas of low correlation, a fine grid (small
$\Delta$) will be required. Figure 2 shows the REM grid size vs the
decorrelation distance of the shadow fading. For a given
interference degradation (say the buffer value plus an additional 2
dB) a large decorrelation distance (say 500 m) enables a coarser
grid size 165 m $\times$ 165 m relative to a decorrelation distance
of 100 m (typical for dense urban areas) when the grid size is 70 m
$\times$ 70 m.

In Fig.~3 we consider a back off in the CR allocation policy. In
order to meet the nominal 2 dB SINR buffer at least 99\% of the
time, the CRs have to target a reduced buffer which is less than 2
dB. Figure 3 shows this buffer vs $\Delta$ for various values of
$D_d$. For a grid size of 25 $\times$ 25 m and a decorrelation
distance of 100 m, the interference buffer is 1.05 dB. Figures 2 and
3 are instructive in determining the grid sizes for different radio
environments.
\subsection{LCR and AED of CR-PU Interference}
Figures 4, 5 and 6 show the LCR (normalized by Doppler frequency) of
the interference for different types of fading and interference
profiles. The interference profiles, i.e., the values of
$I_1,I_2,\ldots,I_N$, are determined for each fading type via a
simulation of the CR allocation policy. From 1000 simulations, two
sets of interferers are selected. The first set has a dominant
interferer and corresponds to the set of interferers with the
highest variance. The second set has no dominant interferer and
corresponds to the set with the least variance.

For all types of fading, the maximum LCR is observed close to the
buffer value. This is because the CR allocation method gives a mean
interference level close to the buffer. Even in strong LOS
conditions ($K=10$ dB), the interference shows a significant number
of level crossings across the buffer due to the scattered component.
Figure~5 shows the case of Rayleigh fading where the interference
budget is dominated by a single large interferer with a number of
smaller additional interferers. Also shown is the case where no
dominant interferer exists. Figure~6 shows the same results for a
Ricean channel with $K=10$ dB. Figures~5 and 6 show that when there
are many small interferers, the resulting interference is more
stable compared to the dominant interferer case. Furthermore, the
Ricean channel always returns a lower LCR as compared to a Rayleigh
channel. The results in Fig.~6 are quite promising. Under the near
LOS conditions that may be present with small cell radii, the CR-PU
interference has a much lower level crossing rate across the
interference buffer for the no dominant interferer case. Hence it
may be a desirable part of the CR allocation policy to avoid any
single user which takes up a significant part of the buffer.
Finally, for completeness, we show the AED results corresponding to
Fig.~6. The AED follows from the LCR using standard results
\cite{Stuber}. As expected the time spent by the interference above
a threshold decreases as the threshold value increases. Therefore,
for the no dominant interferer case, the interference seldom crosses
the threshold (see Fig. 5), and when it does, it only exceeds the
threshold for a small period of time. Finally, we note that all
figures show an excellent agreement between the analytical
approximations and the simulations.

From the point of view of PU system designers, the following
questions are important:
\begin{itemize}
\item How much is the CR-PU interference?
\item Can it be controlled?
\item How often will it exceed a threshold?
\item What happens when it does exceed the threshold? Does it stay
above for a long time or quickly return back to acceptable levels?
\item How do all of the above change with the type of fading?
\end{itemize}
Figures~4-7 shed an interesting perspective on all the above
questions.
\section{Conclusion}
In this paper we have shown that interference degradation to the PU
can be significantly underestimated if the channel state information
needed to estimate interference levels is derived from a coarse REM.
For practical deployments, this may mean that the PU has to accept a
much larger interference from the CRs or the CRs may need to set a
more conservative interference target. This will reduce the number
of CRs allowed. We also determine the LCR and AED for the CR-PU
interference and show that the maximum LCR occurs close to the
maximum allowed interference level for both Rayleigh and Rician
channels. The LCR results show that it is desirable for the
interference to be made up of several small interfering CRs rather
than a dominant source of interference. The LCR of the former case
is more stable than the latter. The AED results also show that the
interference exceed the threshold value for small periods of time in
the latter case.
\section*{Acknowledgment}
The authors wish to acknowledge the financial support provided by
Telecom New Zealand and National ICT Innovation Institute New
Zealand (NZi3) during the course of this research.
\balance
%\nocite{*}
\bibliographystyle{IEEEtran}
\bibliography{IEEEabrv,AusCTW}

\begin{thebibliography}{10}
\providecommand{\url}[1]{#1}
\csname url@rmstyle\endcsname
\providecommand{\newblock}{\relax}
\providecommand{\bibinfo}[2]{#2}
\providecommand\BIBentrySTDinterwordspacing{\spaceskip=0pt\relax}
\providecommand\BIBentryALTinterwordstretchfactor{4}
\providecommand\BIBentryALTinterwordspacing{\spaceskip=\fontdimen2\font plus
\BIBentryALTinterwordstretchfactor\fontdimen3\font minus
  \fontdimen4\font\relax}
\providecommand\BIBforeignlanguage[2]{{%
\expandafter\ifx\csname l@#1\endcsname\relax
\typeout{** WARNING: IEEEtran.bst: No hyphenation pattern has been}%
\typeout{** loaded for the language `#1'. Using the pattern for}%
\typeout{** the default language instead.}%
\else
\language=\csname l@#1\endcsname
\fi
#2}}

\bibitem{report1}
\BIBentryALTinterwordspacing
``Spectrum {P}olicy {T}ask {F}orce {R}eport ({ET} {D}ocket-135),'' {F}ederal
  {C}ommunications {C}omsssion, Tech. Rep., 2002. [Online]. Available:
  \url{http://hraunfoss.fcc.gov/edocs_public/attachmatch/DOC-228542A1.pdf}
\BIBentrySTDinterwordspacing

\bibitem{report2}
M.~A. McHenry, ``{NSF} {S}pectrum {O}ccupancy {M}easurements {P}roject
  {S}ummary,'' {S}hared {S}pectrum {C}ompany, Tech. Rep., 2005.

\bibitem{Mai1}
M.~Vu, N.~Devroye, and V.~Tarokh, ``The primary exclusive regions in cognitive
  networks,'' \emph{IEEE Transactions on Wireless Communications}, April 2008,
  submitted.

\bibitem{ICC}
M.~F. Hanif, M.~Shafi, and P.~J. Smith, ``Interference and deployment issues
  for cognitive radio systems in shadowing environments,'' submitted to the
  \emph{IEEE International Conference on Communications}, 2009.

\bibitem{J.Reed1}
Y.~Zhao, D.~Raymond, C.~da~Silva, J.~H. Reed, and S.~F. Midkiff, ``Performance
  evaluation of radio environment map-enabled cognitive spectrum-sharing
  networks,'' in \emph{Proc. IEEE Military Communications Conference (MILCOM)},
  Oct. 2007, pp. 1--7.

\bibitem{J.Reed2}
Y.~Zhao, L.~Morales, J.~Gaeddert, K.~K. Bae, J.-S. Um, and J.~H. Reed,
  ``Applying radio environment maps to cognitive wireless regional area
  networks,'' in \emph{Proc. IEEE International Symposium on New Frontiers in
  Dynamic Spectrum Access Networks (DySPAN)}, April 2007, pp. 115--118.

\bibitem{Mai}
M.~Vu, S.~Ghassemzadeh, and V.~Tarokh, ``Interference in a cognitive network
  with beacon,'' in \emph{Proc. IEEE Wireless Comm. and Networking Conf.
  (WCNC)}, Apr. 2008, pp. 876--881.

\bibitem{Gud}
M.~Gudmundson, ``Correlation model for shadow fading in mobile radio systems,''
  \emph{IEEE Electronics Letters}, vol.~27, no.~23, pp. 2145--2146, 1991.

\bibitem{Kutz}
N.~L. Johnson, S.~Kutz, and N.~Balakrishnan, \emph{Continuous Univariate
  Distributions, \text{vol. 1}}, 2nd~ed.\hskip 1em plus 0.5em minus 0.4em\relax
  New York: Wiley, 1995.

\bibitem{MRC2}
X.~Dong and N.~C. Beaulieu, ``Average level crossing rate and average fade
  duration of low-order maximal ratio diversity with unbalanced channels,''
  \emph{IEEE Communications Letters}, vol.~6, no.~4, pp. 135--137, July 2002.

\bibitem{MRC1}
P.~Ivanis, D.~Drajic, and B.~Vucetic, ``The second order statistics of maximal
  ratio combining with unbalanced branches,'' \emph{IEEE Communications
  Letters}, vol.~12, no.~7, pp. 508--510, July 2008.

\bibitem{lcrGamma}
R.~Barakat, ``Level-crossing statistics of aperture-integrated isotropic
  speckle,'' \emph{J. Opt. Soc. Amer.}, vol.~5, pp. 1244--1247, 1988.

\bibitem{Kim}
H.-Y. Kim, M.~J. Gribbin, K.~E. Muller, and D.~J. Taylor, ``Analytic,
  computational, and approximate forms for ratios of noncentral and central
  gaussian quadratic forms,'' \emph{Journal of Computational and Graphical
  Statistics}, vol.~15, pp. 443--459, June 2006.

\bibitem{Abromowitz}
M.~Abramowitz and I.~A. Stegun, \emph{Handbook of Mathematical Functions with
  Formulas, Graphs, and Mathematical Tables}.\hskip 1em plus 0.5em minus
  0.4em\relax New York: Dover, 1972.

\bibitem{chiP}
R.~A. Silverman, ``The fluctuation rate of the chi process,'' \emph{IRE Trans.
  Inform. Theory}, vol.~4, no.~1, pp. 30--34, Mar. 1958.

\bibitem{Stuber}
G.~L. Stuber, \emph{Principles of Mobile Communication}, 2nd~ed.\hskip 1em plus
  0.5em minus 0.4em\relax Boston: Kluwer Academic, 2001.

\end{thebibliography}
\end{document}